\documentclass[pre,twocolumn]{revtex4}
\usepackage{epsfig,amssymb,amsmath,latexsym,color,pifont,graphicx,psfrag}
\newcommand{\p}{\varphi}
\newcommand{\om}{\omega}
\newcommand{\er}{\eqref} 
\newcommand{\e}{\varepsilon}

\begin{document}

\title{Emergent multistability and frustration in phase-repulsive networks of oscillators}

\author{Zoran Levnaji\'c}
\affiliation{Department of Physics and Astronomy, University of Potsdam, 14476 Potsdam, Germany}

\begin{abstract}
We study the collective dynamics of oscillator networks with phase-repulsive coupling, considering various network sizes and topologies. The notion of link frustration is introduced to characterize and quantify the network dynamical states. In opposition to widely studied phase-attractive case, the properties of final dynamical states in our model critically depend on the network topology. In particular, each network's total frustration value is intimately related to its topology. Moreover, phase-repulsive networks in general display multiple final frustration states, whose statistical and stability properties are uniquely identifying them.
\end{abstract} 

\maketitle


\section{Introduction} \label{S-introduction}

Complex systems consist of many individual units which generate cooperative functional behavior through mutual interactions~\cite{mikhailov}. In the last decade it has been realized that complex systems can be elegantly described by \textit{networks}, where nodes represent the functional units, and links model their interactions~\cite{d-a-b}. The design and architectures of networks appearing in nature have been extensively studied, revealing a few characteristic classes, such as scalefree or modular networks~\cite{d-a-b,milo,c-f}. A specific emphasis is put on \textit{dynamical networks}, whose collective behavior is a cumulative effect of the individual nodes' dynamics and the underlying network topology~\cite{d-a-b,c-f,arenas}. The interplay between network topology and its emergent dynamics has been widely investigated on various examples of empirical and artificially designed networks. The coherent dynamics on scalefree networks was found to crucially depend on the power-law exponent in the degree distribution~\cite{zhou}. The intra-dependence among the dynamical patterns on micro, meso and macro network scale was investigated~\cite{ja-bosa-teza}. The network constructed by coupling its structural evolution to its emergent dynamics is different from that obtained if these two processes are uncoupled~\cite{diego}. Synthetic gene networks are able to generate various dynamical regimes in relation to the topology of their interactions~\cite{u-a}. Collective effects on natural networks can be examined by assigning the real models as well as formal dynamical systems to the single nodes~\cite{ja-bosa-2}. A recently proposed computational algorithm is able to identify the general dynamical patterns induced by a given topology independently of the particular dynamical process~\cite{jie}.

Models involving \textit{repressive} or \textit{repulsive} interactions play an important role in the context of dynamical networks. The most popular biological examples are the synthetic genetic circuits, in particular \textit{toggle switch} and \textit{repressilator}~\cite{e-g-h}. Genetic circuits consists of a few genes that mutually repress each other creating stable oscillations of their protein concentrations. Dynamical properties of genetic oscillators were extensively studied~\cite{u-a,b-m}. Recent works focus on the systems of interacting genetic oscillators~\cite{u-a,mogens-pablo}, whose cooperative behavior depends on the nature of interactions~\cite{yuan}, which indicates the ways of engineering genetic networks with the desired properties. In addition to genes, sparse repulsive coupling can enhance synchronization in the neural networks~\cite{leyva}. The crucial role of phase-repulsive interactions was studied analytically~\cite{nishikawa} and confirmed experimentally for oscillations in neural astrocyte cultures~\cite{gabor}. Repressive interactions can induce \textit{frustration}~\cite{mogens-pablo}, which in biological systems often generates \textit{multistability}~\cite{angeli}. Existence of multiple operating regimes is essential for biological systems since they provide functional flexibility in responding to the stimuli. This has been largely investigated in relation to genetic oscillators~\cite{u-a,yuan}, with emphasis on biological mechanisms and topological structures leading to multistability~\cite{a-u}. The role of multiple dynamical regimes was also examined in neuronal interactions, both theoretically and experimentally~\cite{newman}.

Rhythmic behavior in many natural phenomena can be described by a \textit{phase variable}~\cite{w-book}, which allows the modeling of complex oscillatory systems and study of the collective effects such as \textit{synchronization}~\cite{prk-book}. The famous Kuramoto model of one-dimensional phase oscillators~\cite{k-book-a} is widely used not only in theoretical studies~\cite{ja-arkady-misha}, but also in modeling specific experimental situations~\cite{ramon}. The phase-attractive coupling model was studied in great detail on a wide range of network sizes and topologies, with various distributions of oscillators' frequencies, involving different coupling schemes and time-delayed interaction~\cite{d-a-b,c-f,arenas,prk-book,k-book-a,ja-arkady-misha}. In general, for sufficiently strong coupling, the system displays a final synchronized dynamical state, which in the case of identical oscillators is always stable full synchronization. Although the time scales of the emergence of synchronization may vary~\cite{arenas}, the final network state is in general independent of the initial conditions. The time-evolution destroys the information on the network structure, since the most ``convenient'' final state always involves synchronization and is often completely unrelated to the underlying topology.

The inherent difference between activatory and repressory interaction is clearly visible in the phase oscillator models. Phase-repulsive oscillators exhibit algebraic relaxation~\cite{daido}, in a sharp contrast with the phase-attractive case. Despite evolving towards zero mean field, arrays of repulsive oscillators display non-trivial dynamical behaviors such as phase locking and clustering~\cite{tsimring-juan}. Networks with a given fraction of repulsive links which induce dynamical frustration were largely studied~\cite{zanette}. In the context of two-dimensional oscillators, the presence of repulsion can improve synchronization~\cite{leyva}, or even generate beam-forming effects that act as a phase array antenna~\cite{rulkov-teresa}. The prescribed synchronization state can be achieved through evolutionary network adaptation by appropriately configuring the repulsive subnetwork~\cite{arizmendi}. Traveling waves were recently found in a globally coupled system with repulsive interactions~\cite{hong}. Among many experimental scenarios, repulsive oscillators were used to model the neuron dynamics with spike timing-dependent plasticity~\cite{yuri} and the cultural dynamics~\cite{kuperman}.

In this paper we consider complex networks of identical oscillators with phase-repulsive coupling. Since the oscillators along each link seek to have the opposite phases, we introduce frustration as a measure of discrepancy with this preferred state for each link. Using average link frustration we characterize the final dynamical states. The frustration--topology relationship is systematically analyzed employing complex networks of various sizes. In opposition to many previous works~\cite{leyva,zanette,rulkov-teresa,arizmendi,hong}, our network model involves \textit{only} phase-repulsive coupling of uniform strength. Considering this simple dynamical model we allow for easier study of the interplay between the emergent dynamics and the underlying topology. As we show, contrary to the phase-attractive models, final state of a network with phase-repulsive coupling is in general frustrated. The network frustration value is intimately related to its topology, and in general increases with its connectivity. Moreover, repulsive networks in general exhibit multiple final dynamical states characterized by different values of link frustration. The structure of the frustration states directly identifies the network topology, suggesting that the repulsive time-evolution preserves much more topological information than the attractive one. Repulsive complex networks thus provide a simple model of the multistable systems.

The paper is organized as follows: in the next Section we introduce our model and examine its basic properties using illustratory examples. In Section \ref{S-sixnode} we systematically study all non-directed networks with six nodes, analyzing the relationship between topology, connectivity and frustration. In Section \ref{S-transitions} we investigate the multistability of a larger network, considering the transitions between frustration states. We discuss our findings and conclude in Section \ref{S-conclusions}.


\section{The model and basic properties} \label{S-themodel}

We consider a network consisting of $N$ oscillators (nodes) with frequencies $\om_i$. Nodes are connected via $L$ non-directed links; $N-1 \le L \le \frac{N(N-1)}{2}$. Dynamical state of the oscillator $i$ is described by the phase variable $\p_i \in [0,2\pi)$, and its dynamics is given by:
\begin{equation}  \dot{\p_i} = \om_i + \frac{\e}{k_i} \sum_{j=1,N}  A_{ij}  g(\p_j - \p_i) \; ,  \label{eq-1} \end{equation}
where $k_i$ is the node's degree ($\sum_i k_i = 2L$), and $\e$ is the coupling strength. Network's topology is expressed through the symmetric adjacency matrix $A_{ij}=A_{ji}$, with value $A_{ij}=1$ if nodes $i$ and $j$ are connected, and $A_{ij}=0$ otherwise. Dynamics starts from a random set of initial phases (IP), selected independently for each oscillator from $\p_i (0) \in [0,2\pi)$. We consider identical oscillators $\om_i = \om$, and take $g=\sin$, thus reducing our system to the simple Kuramoto model. Instead of examining the model with phase-attractive (positive) coupling $\e>0$, we here focus here on the opposite case involving only phase-repulsive (negative) coupling. To this end we fix the coupling strength to $\e=-1$. For simplicity we set $\om = 0$, i.e. put ourselves in the oscillators' rotating reference frame. The equation Eq.\er{eq-1} for our model becomes:
\begin{equation} \dot{\p_i} =  -\frac{1}{k_i} \sum_{j=1,N} A_{ij} \sin (\p_j - \p_i) \; . \label{eq-2} \end{equation}
The interacting pairs of oscillators are seeking to maximize the phase difference between them, i.e. to stretch $\pi$ apart from each other~\cite{daido,tsimring-juan,zanette}. In the final dynamical state, each link will therefore carry the maximal possible phase difference, which is preferably $\pi$. However, as we show in what follows, due to the complex network topology the phase difference along various links if often less than $\pi$, or even zero.

The global dynamical state of an oscillator ensemble is usually quantified via the order parameter $R = \frac{1}{N} \big|  \sum_k e^{i \p_k} \big|$ or one of its variations for complex networks~\cite{arenas}. However, for the purposes of our study, we here resort to a different link-based measure of the collective dynamics. Borrowing the terminology from disordered systems, we define the \textit{frustration} $f_{ij}$ for each link $i-j$ ($A_{ij}=1$) as~\cite{zanette}:
\begin{equation} f_{ij} = 1 + \cos (\p_j - \p_i) \; . \label{eq-f} \end{equation}
Frustration is related with the impossibility of many interacting units to simultaneously attain the state of minimal energy~\cite{mezard}. In our model, a link that stretches to the phase difference $\pi$ has zero frustration, while for a link forced to synchronize ($\p_j-\p_i=0$) the frustration is maximal 2. Frustration measures how ``squeezed'' is a link: it can be pictured as the elastic potential energy contained in it. We characterize the final (stationary) states of dynamical networks Eq.\er{eq-2} by assigning a frustration value $f$ to each link. To measure the global frustration we introduce $F$ as the network average of $f$:
\begin{equation} F = \frac{1}{L} \sum_{i>j} A_{ij} f_{ij} \; , \label{eq-F} \end{equation}
which quantifies how much does the network topology allow for links to stretch. $F$ plays the role of non-equilibrium potential, since Eq.\er{eq-2} can be written as~\cite{zanette}:
\[ \dot{\p_i} =  - \frac{2 L}{k_i} \frac{\partial F}{\partial \p_i} \]
Frustration can be equivalently defined for the phase-attractive coupling, with the preferred link state having zero phase difference. However, since full synchronization is in this case the only final state, all networks will trivially have zero $F$. In contrast, we show here that the topology of a phase-repulsive network is reflected in its final dynamical state.

\begin{figure}[!ht] \centering
\includegraphics[width=0.75\columnwidth]{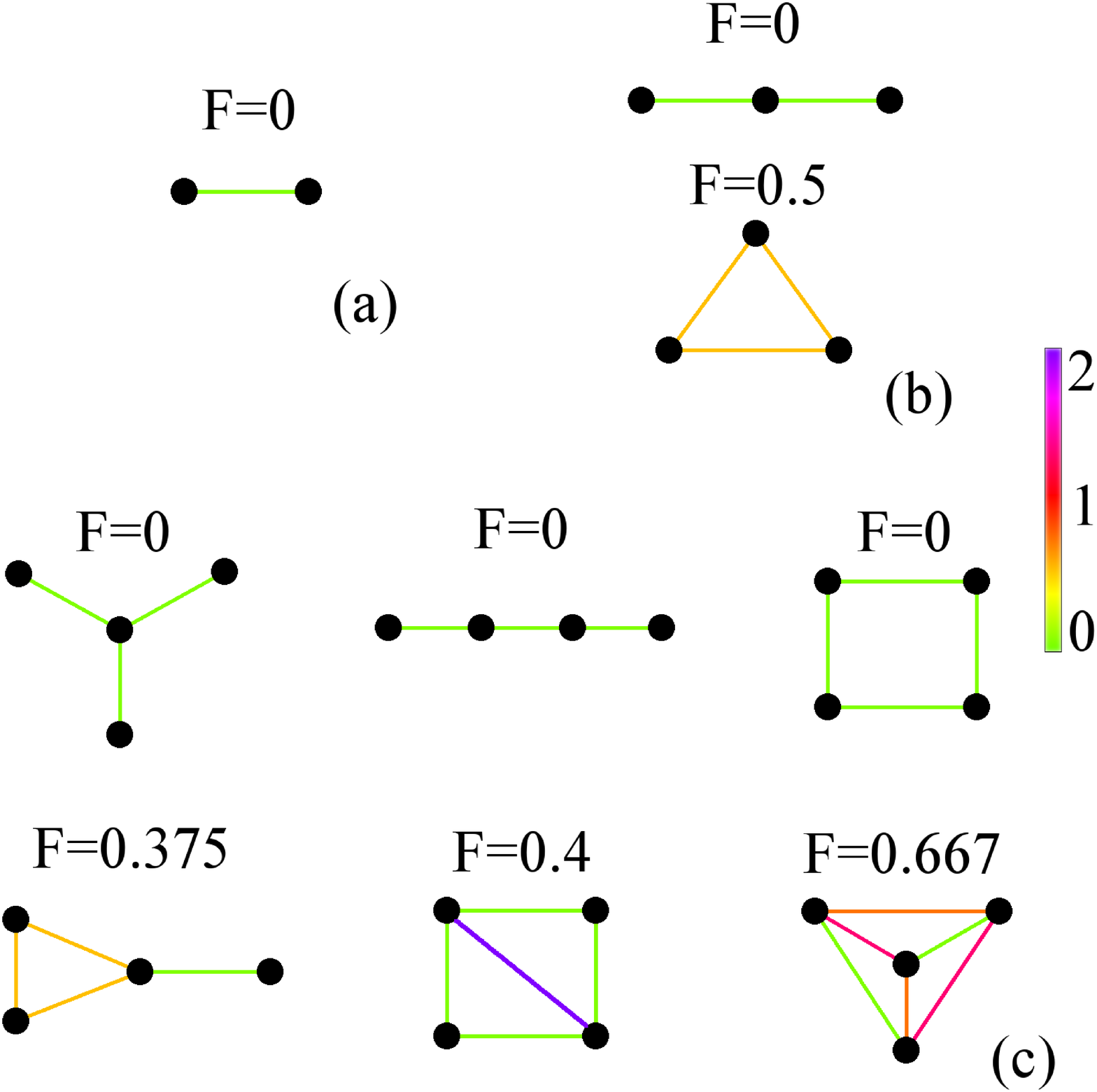} \caption{(color online) Final states of two-node network is (a), both three-node networks in (b), and all six four-node networks in (c). Total frustration $F$ is reported for each network. Links are depicted in a (color)scale that indicates their frustration $f_{ij} \in [0,2]$.} \label{fig-1}
\end{figure}

\textbf{Illustratory examples.} We start with the frustration on small networks. In our visualization of networks, we picture the links using a (color)scale to indicate their frustrations $f_{ij}$. The simplest network of two oscillators shown in Fig.\ref{fig-1}a is never frustrated: its two nodes always attain $\p_1=0$ and $\p_2=\pi$, which yields the phase difference $\pi$ along the link, and hence the frustration $f=0$. Consider a chain of three oscillators (3-chain) shown in Fig.\ref{fig-1}b: if the central node has the phase $\p_2=0$, nothing prevents the other two nodes from having $\p_1=\p_3=\pi$, thus again giving the phase difference $\pi$ and the frustration $f=0$ along each link. The situation is however different in the case of three-node ring (3-ring): since all nodes are now connected, all links can not simultaneously attain the phase difference $\pi$. The stable solution is obtained for $\p_1=0, \p_2=\frac{2\pi}{3}, \p_3=\frac{4\pi}{3}$, i.e. for phase differences $\frac{2\pi}{3}$ (frustration $f=\frac{1}{2}$) along each link. Due to its topology, 3-chain manages to attain a more stretched state than 3-ring.

In Fig.\ref{fig-1}c we show all six four-node networks ordered by increasing $F$. First three of them (top row) achieve $F=0$ due to their specific topologies (we name them 4-star, 4-chain and 4-ring, respectively). Note that 4-ring in opposition to 3-ring attains $F=0$: each diagonal (non connected) pair of nodes is synchronized, yielding $f=0$ along each link. The network with $F=0.375$ can be understood from the discussion of 3-chain and 3-ring above. The fifth network (termed 4-diamond) can be seen as a 4-ring with an additional diagonal link. The outside links manage to stretch to $f=0$ by squeezing the diagonal link to $f=2$, which gives the total of $F=0.4$. Interestingly, this network achieves the minimal frustration by fully squeezing one of its links. The last network is the four-node fully connected graph (4-clique) with $F=\frac{1}{3}$. For different IP this network organizes the values of $f$ differently among the links, always achieving the total of $F=\frac{1}{3}$. The pairs of links that do not share a node have the same $f$-value, and in particular, one such pair is always relaxed to $f=0$, while the other two divide the total of $F=\frac{1}{3}$. The frustration state of 4-clique is thus \textit{degenerate}, with a continuous degeneracy spectrum. The total frustration of four-node networks varies with both topology and number of links. Each of them has a unique way of distributing the frustration among the links: while 4-diamond concentrates it into a single link, other networks distribute it more uniformly.

\begin{figure}[!ht] \centering
\includegraphics[width=0.9\columnwidth]{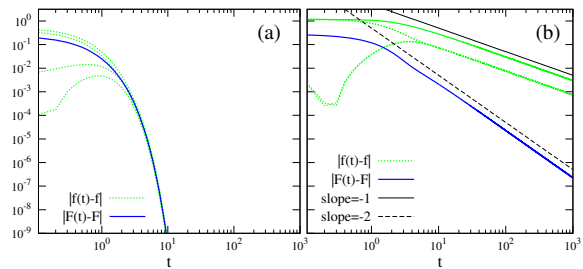} \caption{(color online) Time-evolution of $|f(t) - f|$ and $|F(t) - F|$ for 4-ring in (a), and 4-diamond in (b), for a single IP} \label{fig-2}
\end{figure}

\textbf{Time-evolution.} Contrary to the phase-attractive case, time-evolution of the phase-repulsive networks does not always exhibit exponential relaxation, and directly depends on the network topology~\cite{daido}. To illustrate this, we consider 4-ring and 4-diamond from Fig.\ref{fig-1}c. For each link in each network we examine the behavior of $|f(t) - f|$, where $f$ is the final link frustration value, in addition to $|F(t) - F|$, with $F$ being the final total frustration. We show all the curves for a single IP for 4-ring and 4-diamond in Fig.\ref{fig-2}a and Fig.\ref{fig-2}b, respectively. In the case of 4-ring, all four $f$-values together with the $F$-value display an exponential convergence, similarly to the phase-attractive case. In contrast, all five links of 4-diamond exhibit a power-law convergence with (approximate) slope of $-1$. Interestingly, the convergence of $F$ also shows a power-law, but with a steeper slope of $-2$. 

These two drastically different convergence regimes reflect different dynamical processes: while 4-ring quickly finds its dynamical equilibrium, the diagonal link of 4-diamond resists the phase contraction created by stretching of other four links, thus maintaining the system permanently out of equilibrium. These convergence patterns are robust to IP, and confirm the earlier findings on the relaxation of phase-repulsive oscillators~\cite{daido}. A similar dynamical behavior known as splay states appears in networks of pulse-coupled oscillators~\cite{zillmer}.


\begin{figure*}[!ht] \centering
\includegraphics{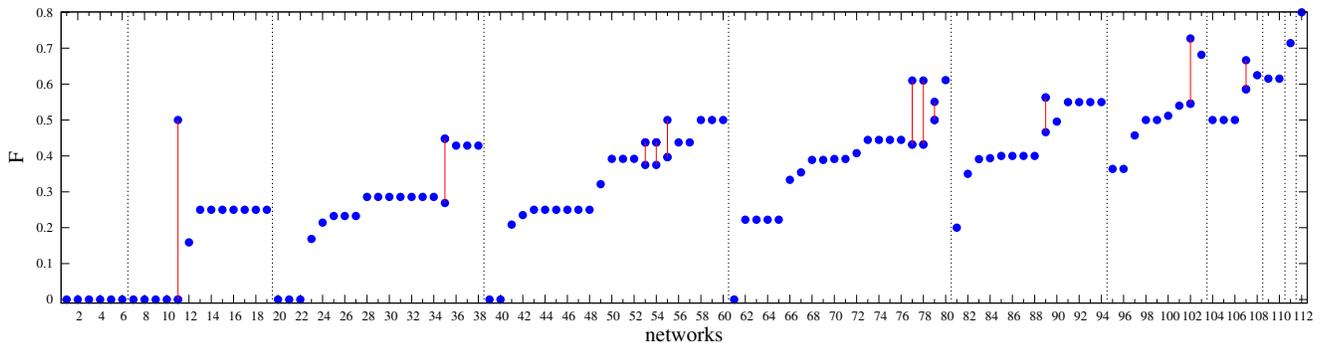} \caption{(color online) Total frustration values $F$ for all 112 connected six-node networks computed for $10^3$ IP. Dashed lines divide between groups with different numbers of links $L$ (networks 1-6 have $L=5$, networks 7-19 have $L=6$ etc.). Within each group networks are order by increasing average $F$ (over IP). 11 networks that exhibit multiple frustration states are marked by the vertical (red) lines.} \label{fig-3}  \end{figure*}

\section{Six-node networks} \label{S-sixnode}

In order to closely examine the relationship between the topology and frustration, in this Section we systematically study the phase-repulsive dynamics of all (connected) six-node networks. There are 112 such networks, which we order by increasing number of links $L$ that range from 5 to 15. For each network we compute the total frustration $F$ for $10^3$ random IP. Results are reported in Fig.\ref{fig-3}, where within each group with the same $L$, we order the networks by increasing $F$ (averaged over IP), thus constructing a numbering of all six-node networks (numbering serves only to identify the networks). Interestingly, there are 11 networks whose final dynamical state can assume two possible values of $F$, depending on the IP (marked by the vertical lines). The remaining 101 networks display a unique frustration state, as do three-node and four-node networks studied previously. The values of $F$ show an overall increase with $L$, finally reaching $F=0.8$ for six-node clique. They also exhibit large variations within each group with the same $L$, which indicates that the total network frustration strongly depends on both topology and $L$. The network group with $L=9$ links (networks 61-80) exhibits the largest variation in frustration values depending on topology; it is also the last group where a fully stretched state with $F=0$ is obtainable. Other groups with medium $L$ (from $L=7$ to $L=10$) display the same trend of topology--frustration relationship, despite containing different number of networks. Some groups also include many networks with similar topologies that all have the same value of $F$.

\begin{figure}[!ht] \centering
\includegraphics[width=0.9\columnwidth]{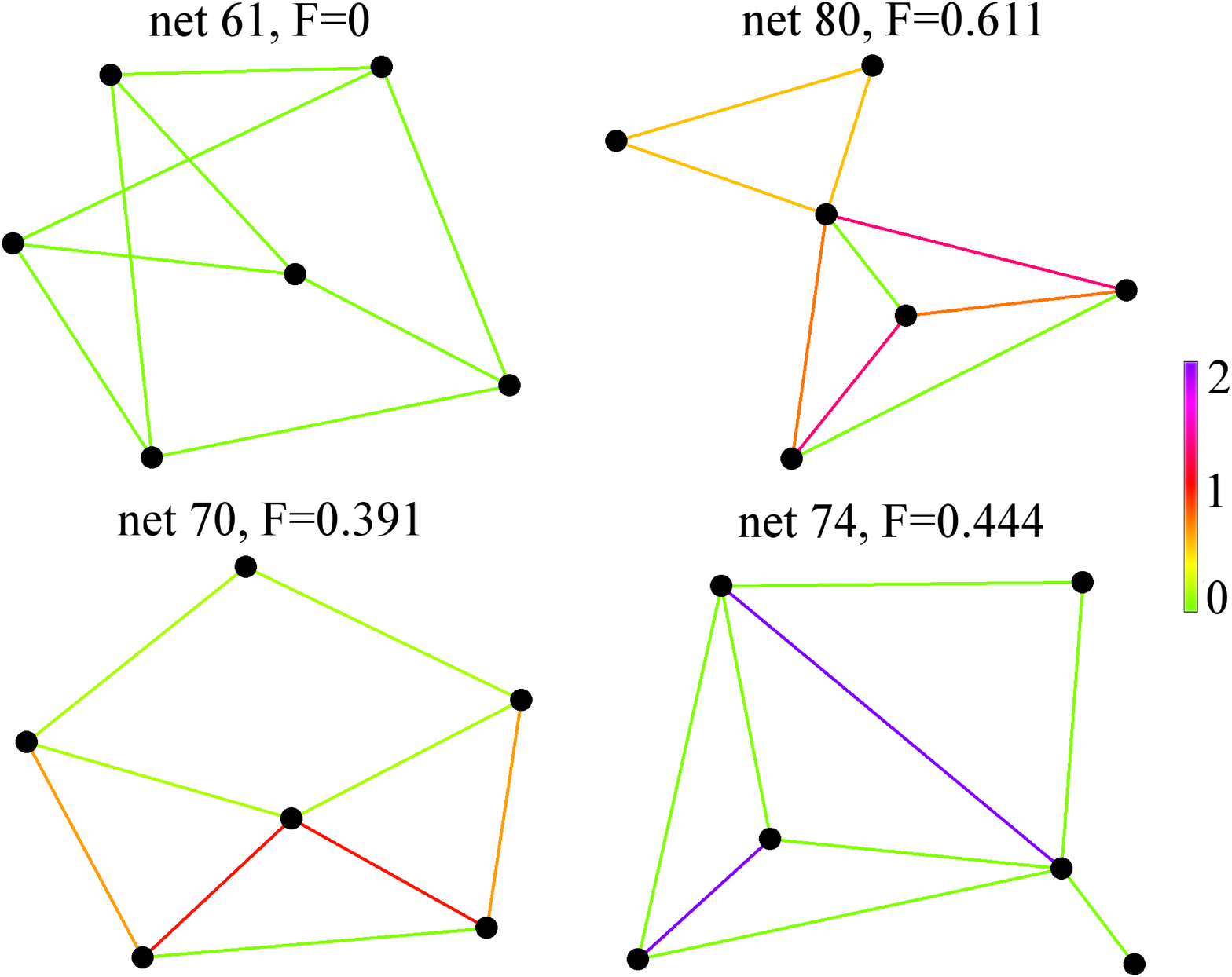} \caption{(color online) Examples of six-node networks displaying a unique frustration state with values of $F$ indicated. Links are marked in (color)scale illustrating their $f$-values.} \label{fig-4} 
\end{figure}

We first examine the networks with a unique frustration state focusing on the $L=9$ group. In Fig.\ref{fig-4} we show the networks 61 and 80 which display minimal and maximal $F$-values in this group, respectively $F=0$ and $F=0.611$. The network 61 manages to stretch to $F=0$ by being constructed from symmetrically organized 4-stars. Network 80, despite having the same $L$, has a much bigger $F$ -- it consists of a 3-ring and 4-clique, both of which have large total frustrations (cf. Fig.\ref{fig-1}). This network exhibits continuously degenerate spectrum of $F$, since it contains the same degeneracy as 4-clique. These two opposite examples testify about the flexibility in containing bigger or smaller frustration within a network, realized through variations of its topology.
In contrast to this, networks 70 and 74 have similar $F$-values despite having rather different topologies. Two networks show very different organization of containing the frustration: while network 70 distributes it over 8 links, network 74 confines it into only 2 links, while completely stretching the other 7. Network 74 includes two 4-diamonds, which in this case display the same frustration pattern as if they were isolated. Each topology has its own way of managing the frustration, that depends on its particularities such as symmetry or modularity.

\begin{figure*}[!ht] \centering
\includegraphics[width=17.0cm]{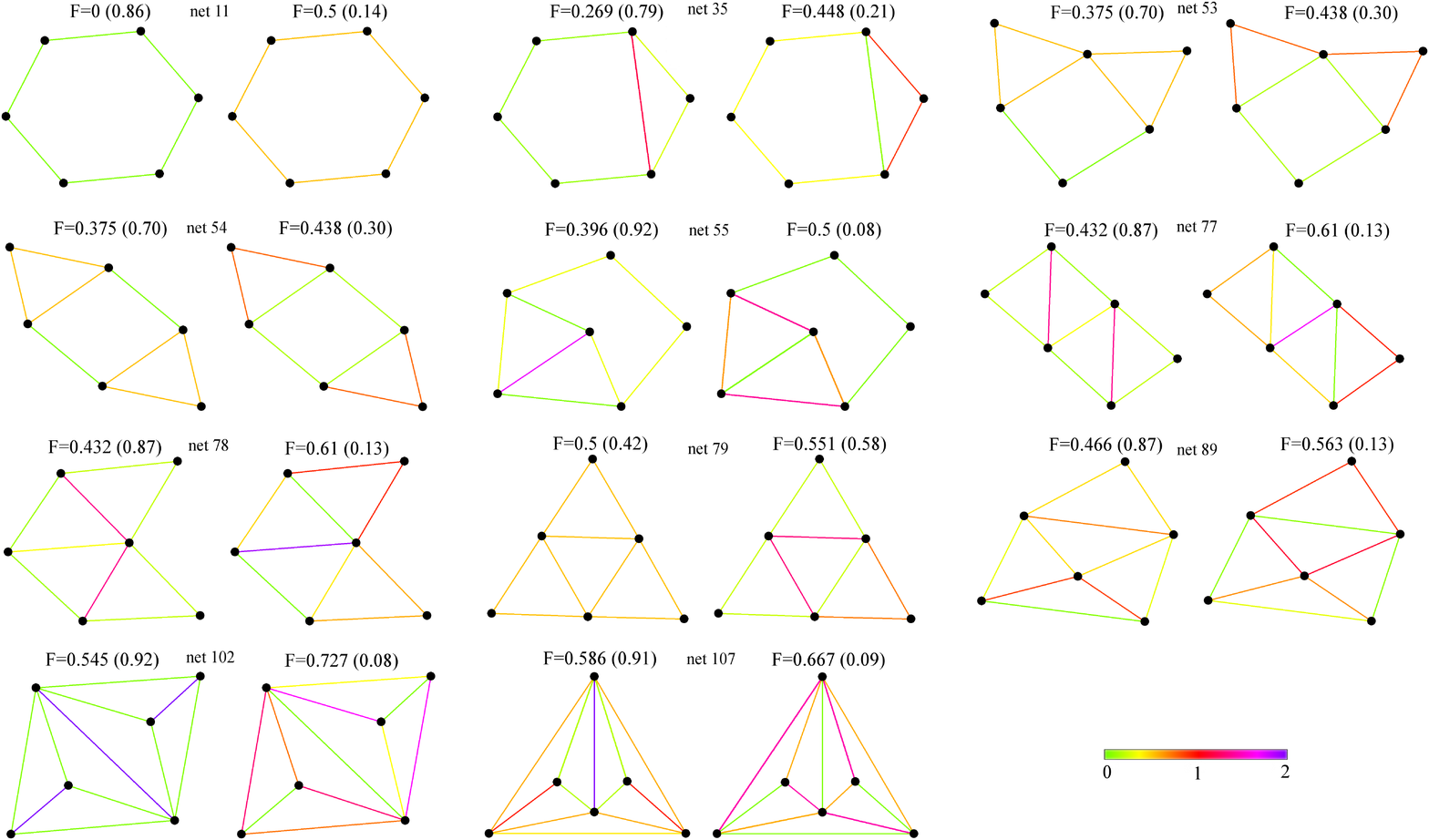}  \caption{(color online) All 11 multistable six-node networks visualized in both frustration states and identified by their numbers (cf. Fig.\ref{fig-3}). The $F$-value is indicated for each state, along with the ratio of IP leading to that state (in parenthesis). (Color)scale illustrates links' $f$-values.} \label{fig-5}
\end{figure*} 

Next we study the examples of networks with multiple frustration states. In Fig.\ref{fig-5} we show all 11 of them, visualized in both states and identified by their numbers as described above (cf. Fig.\ref{fig-3}). For each frustration state we report the $F$-value, along with the fraction of IP leading to it (in parenthesis). There appears to be no specific topological property common to all multiple state networks that would distinguish them from the single state ones. The simplest multistable network 11 (6-ring) can attain the fully stretched state with $F=0$ and a squeezed state with $F=\frac{1}{2}$. The former is obtained for the phase differences $\pi$ for all links (equivalently to 4-ring), while the latter arises for phase differences $\frac{2\pi}{3}$ along each link (equivalently to 3-ring). Both states are stable, but the more squeezed one $F=\frac{1}{2}$ occurs for a smaller fraction of IP (only 14\%). This is to say that both states are stable fixed points (sinks) for the dynamical system Eq.\er{eq-2} with 6-ring topology, but the $F=0$ state has a bigger basin of attraction. Network 35 is a 6-ring with an additional link inside: two frustration states differ in distributing the frustration between these two subnetworks. Networks 53 and 54 are topologically similar, and consequently have the same $F$-values occurring for the same fractions of IP. Their dynamics is a competition between a 4-ring and two 3-rings in escaping the frustration. The $F=0.396$ state of network 55 exhibits a discrete degeneracy: opposite pairs of links in 4-ring can swap their $f$-values without changing $F$, while the $F=0.5$ state shows a continuous degeneracy, similar to 4-clique. The remaining multistable networks display similar patterns: the difference between two states generally lies in the competition between two network's structural elements in escaping the frustration by attempting to stretch to the maximal attainable phase difference. The choice of IP pre-defines the final state. The states with lower $F$-values are usually more preferred. The exception to this is network 79: its higher $F$ state appears more frequently, despite the lower $F$ state involving a uniform distribution of $f$-values over all links. Two $F$-values are typically close, although not always (network 11). Additional stable states are in principle possible for some networks, but with extremely small basins of attraction, which makes them very difficult to observe. Each of the above networks was tested for $10^4$ IP, and no third stable state was found.

Relaxation of the six-node networks displays the same exponential and power-law convergence patterns observed earlier (Fig.\ref{fig-2}). Power-law relaxation, testifying about the non-equilibrium processes on the network, is typically found on the network such as 74 (Fig.\ref{fig-4}), which concentrate their entire frustration into a few completely squeezed links ($f=2$). Interestingly, we revealed various power-law slopes for some links in those networks, that indicate different squeezing strengths exerted by the rest of the network, which is a direct consequence of their specific topologies~\cite{daido}.

\begin{figure}[!ht] \centering
\includegraphics[width=0.9\columnwidth]{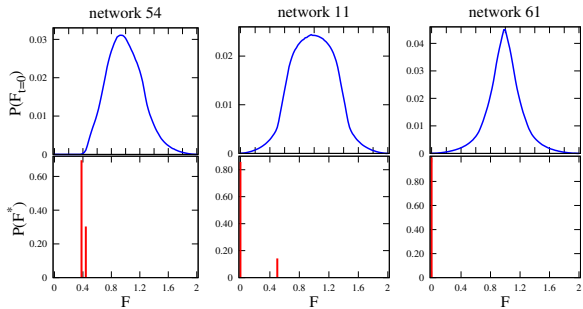} \caption{(color online) Distributions of initial (top panel), and final (bottom panel) values of $F$ for many IP, for networks 54, 11 and 61 (cf. Figs.\ref{fig-4}\&\ref{fig-5}).} \label{fig-6}
\end{figure}

It is instructive to consider the distributions (over IP) of initial total frustration $F$ for various networks, since they reflect their topological symmetries. In Fig.\ref{fig-6} (top panel) we show the distributions of $F$-values at time $t=0$ for networks 54, 11 and 61 (cf. Figs.\ref{fig-4}\&\ref{fig-5}). The central symmetry of distributions for networks 11 and 61 indicates that each link has the same structural ``role'' in relation to the phase-repulsive dynamics. 
As expected, all links of those networks always have the same final values of $f$. On the other hand, the distribution for network 54 is asymmetric, since not all of its links ``see'' the network in the same way. In the bottom panel of Fig.\ref{fig-6} we show the distributions of final $F$-values, which for networks 54 and 61 consist of two possible values in a given ratio, and for network 61 a unique value $F=0$. Note that for all networks, the lowest final frustration state is also the state with the lowest possible frustration -- e.g., for network 54, no situation with $F$ smaller than $F=0.375$ is obtainable due to its topology.

\begin{figure}[!ht] \centering
\includegraphics[width=0.9\columnwidth]{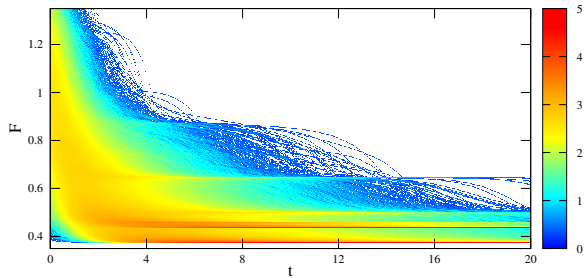} \caption{(color online) Time-evolution of the distribution of $F$ for network 54: vertical coordinate indicates the fraction of IP (log-scale) with a certain value of $F$ at time $t$. $10^5$ IP were considered for $t \in [0,20]$.} \label{fig-7}
\end{figure}

In Fig.\ref{fig-7} we show the time-evolution of $F(t)$ for $10^5$ IP for network 54. The vertical coordinate indicates the fraction of IP having a certain value of $F$ at time $t$ (time from $t=0$ to $t=20$ is considered). 
We examine the evolution of the initial distribution of $F$ into its two final states (cf. Fig.\ref{fig-6} left side). Many intermediate unstable states with higher $F$ are visited during the evolution before settling into one of the final states. For instance, a state with $F \approx 0.65$ persists for some time, but eventually decays (network 54 was tested for $10^6$ IP and no third stable state was revealed). This cascading dynamics involving higher frustration states also occurs in single state networks. Phase-repulsive networks almost always display many possible final states, whose stability however crucially depends on their topology. The appearance of multistability can be seen as the persistence of states with higher $F$ due to the topological details.

Smaller networks (of size $N<6$) do not exhibit multiple frustration states. On the other hand, multistability becomes common as the network size is increased: many networks of size $N=7$ are multistable, some of them possessing three states. As we show in two remaining Sections, the number and the organizational complexity of frustration states dramatically increases with the network size and complexity.


\begin{figure*}[!ht] \centering
\includegraphics[width=17.0cm]{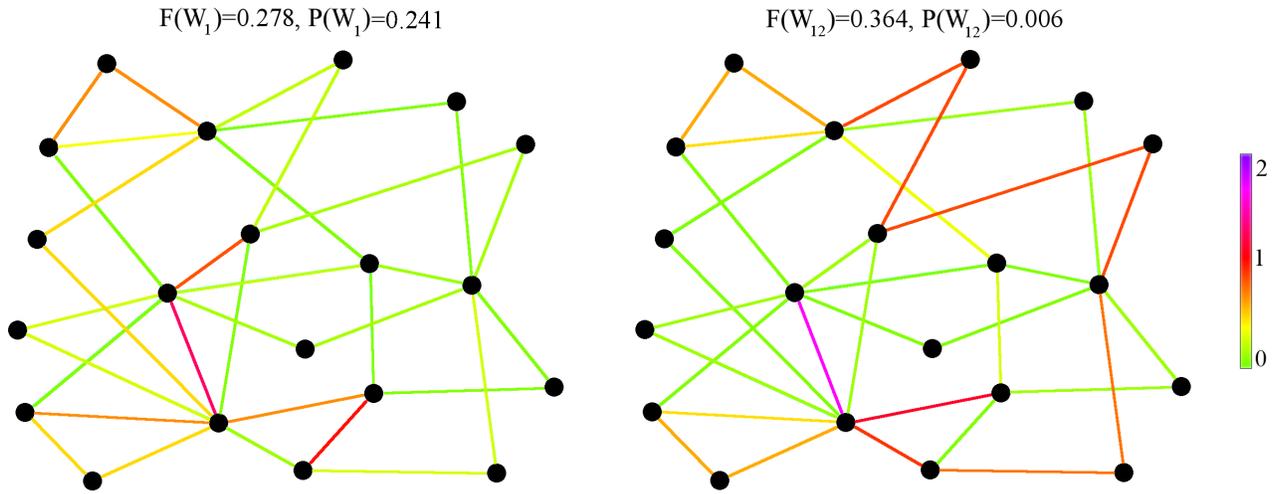}  \caption{(color online) Lowest (left) and highest (right) frustration state for 20-node network, $W_1$ and $W_{12}$, visualized with (color)scale indicating $f$. $F$-values and $P$-values are reported.} \label{fig-8}
\end{figure*} 

\section{Frustration states on a large network} \label{S-transitions}

In this Section we examine a larger network with more frustration states, and study the transitions among them occurring by perturbing the dynamics. To this end, we construct a network with $N=20$ nodes as follows: starting from 3 initially unconnected nodes, we add at each step one new node to the existing network. Each new node is preferentially attached to two existing nodes, randomly chosen with probabilities proportional to $k_i + \alpha$, where $k_i$-s are the current node degrees, and $\alpha=1.1$. The described step is repeated 17 times until the network size of $N=20$ is reached, resulting in a 20-node network with $L=34$ links. Phase-repulsive dynamics Eq.\er{eq-2} is implemented as above.

\begin{figure}[!ht] \centering
\includegraphics[width=0.9\columnwidth]{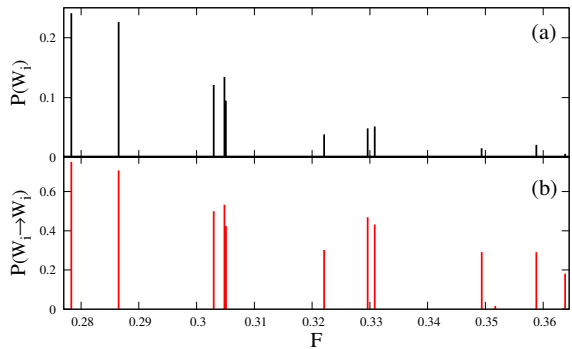} \caption{(color online) 12 frustration states $W_1, W_2, \hdots W_{12}$ for 20-node network shown through their values of $F(W_i)$. (a): fraction of IP $P(W_i)$ leading to each state $W_i$. (b): the fraction of random kicks leading to no change in the state $P(W_i \rightarrow W_i)$  (cf. Fig.\ref{fig-12}b).} \label{fig-9}
\end{figure}

\begin{figure}[!ht] \centering
\includegraphics[width=0.9\columnwidth]{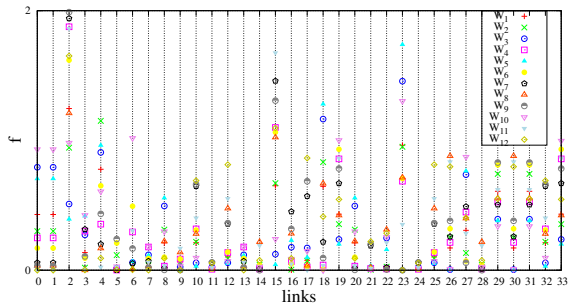} \caption{(color online) Link frustration values $f$ for all links and all states of 20-node network (arbitrary numeration). Different symbols are used for different states $W_i$ (legend).} \label{fig-10}
\end{figure}

The dynamics on this preferential attachment grown network displays twelve final frustration states. The network is visualized in Fig.\ref{fig-8} in its lowest (left) and highest (right) frustration state. We name the states $W_1, W_2, \hdots W_{12}$, indexing them by increasing $F$-value termed $F(W_i)$. Each state $W_i$ appears for a certain fraction of IP which called $P(W_i)$. All 12 values of $P(W_i)$ are reported in Fig.\ref{fig-9}a in relation to the corresponding $F(W_i)$. The most preferred state $W_1$ is also the one with the lowest $F(W_1)=0.278$. The values of $P(W_i)$ overall decrease with $F(W_i)$, although the least preferred state is $W_{10}$ with $P(W_{10}) \lesssim 10^{-3}$ and $F(W_{10})=0.352$, while for the highest frustration state $W_{12}$ we find $F(W_{12})=0.364$ and $P(W_{12})=0.006$. 
The states are very unequally spaced in $F$, with $F(W_4)$ and $F(W_5)$ being nearly the same. Each state can be characterized by its specific distribution of link frustrations $f$. As shown in Fig.\ref{fig-8}, in state $W_1$ the network stretches most of the peripheral nodes, and confines the entire frustration into the links between hubs. In contrast, network in $W_{12}$ stretches most of the links around the central hub, while squeezing some of the outer links. Similarly, the differences between other states typically relate to dividing the frustration between the central and peripheral links. In Fig.\ref{fig-10} we report all $f$-values for all links and all states (numeration of links is arbitrary and serves only to discern among them). 
Some links (e.g. 2, 15, 28, 23) exhibit a wide range of attainable $f$-values depending on $W_i$, that covers the entire $[0,2]$ interval. Other links (e.g. 7, 11, 13, 24) always maintain roughly the same $f$-value regardless of $W_i$. The former group of links is flexible to different dynamical situations, while the latter group is robust to it. Some links such as 21 even exhibit two groups of $f$-values. Some pairs of links always have the same $f$-values for all states $W_i$, which suggests that they have the same dynamical role in the network (e.g. 0 and 1, 8 and 20, 12 and 25, 26 and 30), as also visible in Fig.\ref{fig-8}. The network's response to phase-repulsive dynamics involves different dynamical roles for different links, realized through a spectrum of link frustrations and their flexibility.

\begin{figure}[!ht] \centering
\includegraphics[width=0.9\columnwidth]{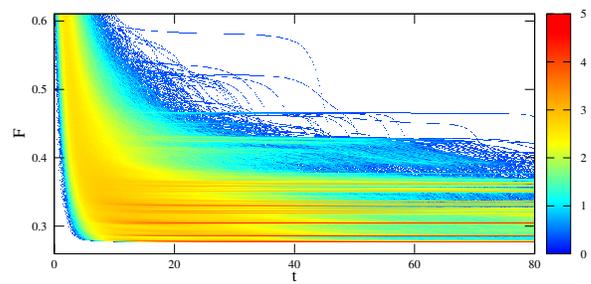} \caption{(color online) Time-evolution of initial $F$ distribution for 20-node network, as done in Fig.\ref{fig-7} for network 54. $10^5$ IP were considered for $t \in [0,80]$.} \label{fig-11}
\end{figure}

In Fig.\ref{fig-11} we illustrate the time-evolution of the initial distribution of $F$, as done previously for network 54 (cf. Fig.\ref{fig-7}). The system now visits an even larger number of intermediate unstable states with the higher $F$ prior to settling in one of the $W_i$. 
The speed of this cascading process seems to decrease with time. The complex topology of the underlying network is clearly reflected in the complexity of general time-evolution. The values of $P(W_i)$ are a consequence of starting the dynamics from random IP, which yields the initial $F$-value much bigger than the range of $F(W_i)$. Starting the dynamics from specific IP will not influence $W_i$, but it will change $P(W_i)$.

\begin{figure}[!ht] \centering
\includegraphics[width=0.9\columnwidth]{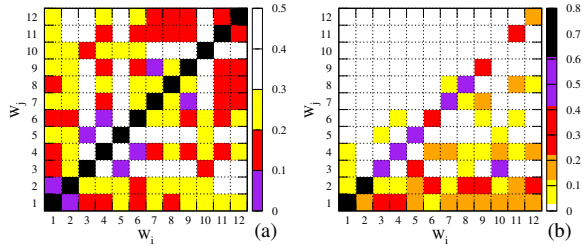} \caption{(color online) (a): Hamming distances between frustration states $W_i$ and $W_j$ for 20-node network. (b): transition rates $P(W_i \rightarrow W_j)$ defined as fractions of random kicks yielding a given transition. Note the piecewise constant (color)scale.} \label{fig-12}
\end{figure}

We further compute the Hamming distance $H(W_k,W_l)$ between the states defined as:
\[ H(W_k,W_l) =  \frac{1}{L} \sum_{i>j}  A_{ij} \big|f_{ij}(W_k) - f_{ij}(W_l) \big| \; , \]
which quantifies the ``frustration distance'' between any two states by averaging the difference in link frustrations over all links. The symmetric matrix of Hamming distances for the 20-node network is shown in Fig.\ref{fig-12}a.
The states can roughly be divided into four clusters:
\begin{itemize}
 \item $W_1$ and $W_2$ have similar distances to all other states, and are mutually very close.
 \item the same holds for $W_3$, $W_5$ and $W_{10}$, which are also close to $W_1$ and $W_2$.  
 \item the cluster of states $W_7, W_8, W_9, W_{11}, W_{12}$ are higher $F$ states that are mutually relatively close, but far from the second, and somewhat close to the first cluster.
 \item $W_4$ and $W_6$ are again mutually very close, and also close to the first and the third cluster, while far from the second.
\end{itemize}
The clustering of frustration states is another property of phase-repulsive dynamics that reflects the network topological details. Note that this classification seems not to be directly correlated with the values of $F(W_i)$ 
and $P(W_i)$: for instance, states $W_4$ and $W_5$ are far from each other despite having almost the same $F$-values.

Below we investigate the transitions between the frustration states on the 20-node network induced by the random perturbations of the network dynamics. To this end we modify the Eq.\er{eq-2} by adding the kick term:
\[ \dot{\p_i} = - \frac{1}{k_i} \sum_{j=1,N} A_{ij} \sin (\p_j - \p_i)  \, + \, K_i \sin (\p_i + \alpha_i) \delta(t-T) \]
which acts at time $t=T$ by independently perturbing the dynamics of each node~\cite{ja-arkady-misha}. For each kick and each node, we randomly choose the kicking strength $K_i$ from a Gaussian distribution centered at zero with standard deviation 2, and the phase-shifts $\alpha_i$ uniformly from $[0,2\pi)$. The network is prepared at time $t=T$ in state $W_i$, after which the kick is applied. Upon perturbation, the network settles into a new state $W_j$. This procedure is repeated $10^3$ times for each starting state $W_i$, and the transitions $W_i \rightarrow W_j$ are recorded. We denote with $P(W_i \rightarrow W_j)$ the fraction of random kicks leading from the state $W_i$ to the state $W_j$. The matrix of transitions is reported in Fig.\ref{fig-12}b, where the scale shows the values of $P(W_i \rightarrow W_j)$. The matrix is of course non-symmetric, since it is easier to induce the transitions from a higher to a lower $F$ state than vice versa. Similarly, the transitions from a more preferred into a less preferred state are more common than the inverse transitions. In general, each state $W_i$ appears to have more and less preferred states $W_j$ into which it jumps. The obtained transition rates seem to reflect the clustering of the states according to the matrix of Hamming distances shown in Fig.\ref{fig-12}a: the high $F$ states prefer to jump into lower $F$ states that belong to their own cluster. For instance, while all states jump into $W_1$ and $W_2$ (with various ratios), only some states jump into $W_3$, such as members of its cluster $W_5$ and $W_{10}$. Interestingly, among the very few transitions occurring from a lower into a higher $F$ state, most start from $W_1$, while all others occur within a given cluster.

A special role is played by the transitions that do not change the frustration state, i.e. $W_i \rightarrow W_i$, which are the diagonal elements of the matrix $P(W_i \rightarrow W_j)$ in Fig.\ref{fig-12}b. They are indicators of the robustness of a given state against the perturbations. We show the values of $P(W_i \rightarrow W_i)$ in Fig.\ref{fig-9}b for comparison with the corresponding $P(W_i)$ in Fig.\ref{fig-9}a. The ratios of $P(W_i \rightarrow W_i)$-values only partially reflect the ratios of $P(W_i)$-values: while $W_1$ and $W_2$ are the most robust states, higher $F$ states are more robust than expected, in particular $W_7, W_8, W_9$ and $W_{11}$. The value of $P(W_i)$ -- fraction of IP leading to $W_i$ (basin of attraction), can be seen as the ``width'' of the potential hole defining $W_i$. Similarly, the value $P(W_i \rightarrow W_i)$ can be understood as the ``depth'' of the potential hole, as it indicates how strong perturbation is needed to jump out of $W_i$. The comparison of Fig.\ref{fig-9}a and Fig.\ref{fig-9}b reveals that depths and widths of the states are not completely correlated: low $F$ states are wide and relatively deep, while many high $F$ states are only somewhat shallower despite being much narrower. Note that the selection of the kicking strengths is done appropriately to allow for these properties to be observed. Very strong perturbations would erase the memory of starting state $W_i$, and all transitions would follow the same probabilities as if starting from random IP.

\begin{figure}[!ht] \centering
\includegraphics[width=0.9\columnwidth]{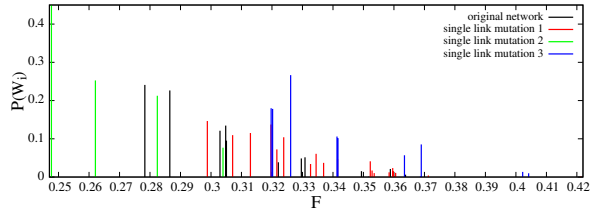} \caption{(color online) 12 frustration states for original 20-node network (in black, same as Fig.\ref{fig-9}a), along with three different profiles obtained for three examples of mutated network (see text for details).} \label{fig-13}
\end{figure}

Finally, we examine the uniqueness of the network frustration profile (shown in Fig.\ref{fig-9}a) in relation to the network topology. We implement the link mutation scheme as follows: one node of the original network and one of its links are chosen at random. The link is then re-wired to a different (randomly chosen) node, making sure that the network stays connected. The resulting network differs from the original one only in a single link, so it is still ``topologically close'' to it. We compute the statistics of $F(W_i)$ and $P(W_i)$ for many mutation examples: the profile always drastically differs from the original profile from Fig.\ref{fig-9}a. To illustrate this, we show in Fig.\ref{fig-13} the original profile (black), together with three examples of profiles obtained for networks with a single link mutation. 
The first of them has 20 states, while the second one has only 4; the third profile has the most preferred state different from the lowest $F$ one. It appears that even a single link mutation, which only marginally changes the topology, yields a dramatic change in the number and the properties of the frustration states. This extreme sensitivity of multistability to the topology again testifies about the intricate relationship between them; it appears that the frustration profile is in general unique for large networks. This can facilitate the reconstruction of the phase-repulsive networks from the dynamical data, in which context various methods are already in use~\cite{ja-arkady-recon}.


\section{Discussion and Conclusions}  \label{S-conclusions}  

We studied the collective dynamics of identical Kuramoto oscillators with phase-repulsive interactions on non-directed complex networks. Various network sizes and topologies were considered: all 112 connected six-node networks were systematically examined, in addition to a preferential attachment grown 20-node network. In opposition to the phase-attractive case, our model involves dynamical frustration resulting from the tendency of linked oscillator pairs to attain the maximal difference of $\pi$ between them, which is not always possible due to the network's topological complexity. We showed that each network has its characteristic total frustration $F$, which largely depends on its size and topology. Moreover, certain networks display multiple frustration states in relation to different initial conditions, which can be classified into clusters. Transitions between states also reflect topological details and cluster organization. As we finally showed, the profile of frustration states appears to be a unique ``fingerprint'' for each network, which is associated with methods of detecting the network structure from dynamical data~\cite{ja-arkady-recon}.

In the presence of noise our model is expected to exhibit less frustration states; shallow states such as $W_{10}$ in 20-node network will immediately become unstable. With increase of noise strength more states will lose stability, finally reaching the point where only a single state will remain accessible. This state will thus be the unique final dynamical state for a phase-repulsive network.

\begin{figure}[!ht] \centering
\includegraphics[width=0.9\columnwidth]{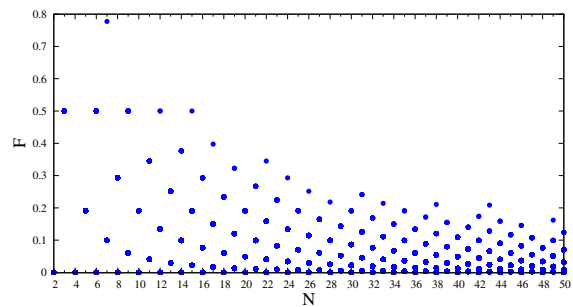} \caption{(color online) Total frustration $F$ for all rings with $N=2,\hdots 50$ nodes computed for $10^3$ IP.} \label{fig-14}
\end{figure}

For networks with the ring topology the $F$ value is directly related to the ring's parity. Recall from Fig.\ref{fig-1} that 2-node network and 4-ring have $F=0$, while 3-ring has $F=0.5$. To systematically study this, we show in Fig.\ref{fig-14} the $F$-values of rings as function of number of nodes $N$. Rings with even $N$ always have the lowest $F$ value $F=0$; on the other hand, rings with odd $N$ are always frustrated, but their lowest $F$-value approaches zero.
With increase of $N$, rings display a growing number of multiple frustration states, starting with 6-ring (cf. Fig.\ref{fig-5}). Multiple $F$-values exhibit a periodic pattern with $N$. The range of attainable $F$-values shrinks with increase of $N$, approaching zero at the limit $N \leftarrow \infty$, where the ring topology approaches that of a chain. The relationship between frustration and parity is associated with the methods of detecting network \textit{motifs} -- overrepresented subnetworks with specific topologies~\cite{d-a-b,milo}. Some methods of searching for network rings are already in use~\cite{mayan}. Since link frustration $f$ contains local network information, it could be in principle used for motif detection. However, this will crucially depend on the way motif is embedded in the network -- both 4-diamond and 4-clique (cf. Fig.\ref{fig-1}) contain 3-ring as motif, but its $f$-values are different from those found on isolated 3-ring. One could also seek to generalize the idea of parity in the context of networks and find a common topological property for all networks with $F=0$.

\begin{figure}[!ht] \centering
\includegraphics[width=0.9\columnwidth]{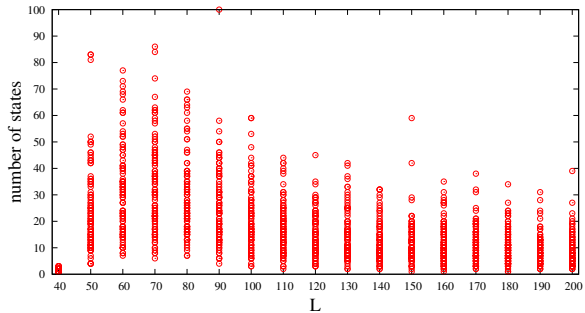} \caption{(color online) Number of observed $F$ states for each realization of Erd\H os-R\'enyi random graph with $N=40$ nodes and $L$ links between $L=40$ and $L=200$. 200 IP were considered for each realization.} \label{fig-15}
\end{figure}

Another immediate question revolves around the number of frustration states in relation to the network connectivity (number of links $L$). To examine this we construct Erd\H os-R\'enyi random graphs with $N=40$ nodes, taking multiples of 10 for $L$ between $L=40$ and $L=200$ ($L=40,50,\hdots 200$)~\cite{d-a-b,c-f}. For each $L$-value, we construct 100 different random graph realizations, and record the total number of observed states after 200 runs (phase-repulsive dynamics is implemented as previously). The results are shown in Fig.\ref{fig-15}: biggest numbers of states ($\gtrsim 50$) most often occur on sparse networks around $L \sim 70$. 
Sparse networks also exhibit the largest range of possible number of states depending on the topology, and seem always to have no less than 5 states. With increase of $L$, networks typically display between 1 and 30 states, which does not substantially change even for very big $L$. On the other hand, too sparse networks have even less states. This result might relate to the sparse connectivity observed in many biological and technological networks~\cite{d-a-b,c-f}, and emphasize the dynamical properties of sparse topologies. This also indicates the optimal range of network connectivity for modeling complex multistable systems. The 20-node network studied in Section \ref{S-transitions} is also sparse.

A further question regards the design of networks with minimal or maximal total frustration. We showed in Figs.\ref{fig-3}\&\ref{fig-4} that a network with a fixed number of links may have very different $F$-values depending on its topology. It would be interesting to examine the topological differences between large networks with fixed $L$ having minimal and maximal $F$. Picturing $F$ as an elastic potential energy contained in the network, this model may indicate the design algorithms for construction of maximally squeezed (or stretched) elastic networks. Similar question refers to the networks with minimal or maximal number of states, which might be of interest in modeling multistable complex systems (cf. Fig.\ref{fig-15}).

Future generalizations include networks with non-identical oscillators, which are expected to exhibit an even wider spectrum of frustration states, including multirhythmicity~\cite{u-a}. The interaction function $g$ from Eq.\er{eq-1} was here taken $g=\sin$, although other choices of odd $g$ might be interesting. Repulsive dynamics on directed and weighted networks is still poorly understood. The stability of the fixed points of dynamical system Eq.\er{eq-2} can also be investigated analytically, using the network Laplacian defined as $L_{ij} = k_i \delta_{ij} - A_{ij}$~\cite{d-a-b,c-f,arenas}. Drawing conclusions about network multistability by examining $L_{ij}$ might allow more detailed and systematic insights. In particular, it would be interesting to study the properties of the Laplacian eigenvalues in relation to the network relaxation patterns (cf. Fig.\ref{fig-2}).


\acknowledgments Thanks to A.~Pikovsky and A.~D\'iaz-Guilera for useful suggestions. Special thanks to Agentur f\"ur Arbeit Potsdam.

\end{document}